\documentclass[twocolumn]{aastex63}

\newcommand{\beq}{\begin{equation}}
\newcommand{\eeq}{\end{equation}}
\newcommand{\beqn}{\begin{eqnarray}}
\newcommand{\eeqn}{\end{eqnarray}}
\newcommand{\gppr}{\stackrel{>}{\scriptstyle \sim}}

\shorttitle{Particle Acceleration in Shearing Flows}
\shortauthors{Rieger and Duffy}

\begin{document}

\title{Particle Acceleration in Shearing Flows: Efficiencies and Limits}



\correspondingauthor{Frank M. Rieger}
\email{f.rieger@uni-heidelberg.de}

\author{Frank M. Rieger}
\affiliation{ZAH, Institute of Theoretical Astrophysics, University of Heidelberg \\
Philosophenweg 12, 69120 Heidelberg, Germany}
\affiliation{Max-Planck-Institut f\"ur Kernphysik, Saupfercheckweg 1, 69117 Heidelberg, Germany}

\author{Peter Duffy}
\affiliation{School of Physics, University College Dublin, Belfield, Dublin 4, Ireland}

\nocollaboration{2}



\begin{abstract}
We examine limits to the efficiency for particles acceleration in shearing flows, showing that 
relativistic flow speeds are required for efficient gradual shear acceleration. We estimate 
maximum achievable particle energies for parameters applicable to relativistic AGN jets. 
The implications of our estimates is that if large-scale jets are relativistic, then efficient 
electron acceleration up to several PeV, and proton acceleration up to several EeV energies 
appears feasible. This suggests that shear particle acceleration could lead to a continued 
energization of synchrotron X-ray emitting electrons, and be of relevance for the production 
of ultra-high-energy cosmic-ray particles.
\end{abstract}


\section{Introduction}
Shear flows are expected to be present in various astrophysical environments. Prototypical examples 
include black-hole accretion flows and the relativistic outflows or jets in gamma-ray bursts and Active Galactic 
Nuclei (AGN) \citep{Rieger2004}. The jets in AGN, for example, are likely to exhibit some internal velocity 
stratification from the outset, shaped by a highly relativistic, ergo-spheric driven (electron-positron) flow 
that is surrounded by a slower moving (electron-proton dominated) wind from the inner parts of the disk 
\citep[e.g.,][and references therein]{Marti2019,Fendt2019}. As these jets propagate, interactions with the 
ambient medium is know to excite instabilities and to induce mass loading, enforcing further velocity shearing 
\citep[e.g.,][and references therein]{Perucho2019}. Radio images of parcec-scale jets in AGN indeed provide 
phenomenological evidence for internal jet stratification, examples including limb-brightened structures or 
boundary layers with parallel magnetic fields  \citep[e.g.,][]{Giroletti2008,Blasi2013,Piner2014,Nagai2014,
Gabuzda2014,Boccardi2016}. When taken together, this suggests that transversal velocity stratification and 
shear is a generic feature of AGN-type jets. Given the diversity of observed emission properties, this has 
in recent times generated new interest in multi-zone or spine-shear-layer acceleration and emission models 
\citep[e.g.,][]{Sahayanathan2009,Laing2014,Tavecchio2015,Rieger2016,Liang2017,Chhotray2017,Liu2017,
Kimura2018,Webb2018}.\\ 

Shear flows can in principle facilitate particle acceleration by several means \citep[see][for recent review]{Rieger2019}. 
One prominent possibility includes a stochastic Fermi-type mechanism, in which particle energization occurs as 
a result of elastically scattering off differentially moving (magnetic) inhomogeneities \citep[e.g.,][]{Berezhko1981,
Earl1988,Webb1989,Rieger2006,Lemoine2019}. In gradual shear particle acceleration these inhomogeneities are 
considered to be frozen into a background flow whose bulk velocity varies smoothly in the transverse direction. 
The scattering center's speeds are thus essentially characterised by the general bulk flow profile. Given recent 
developments, the present paper studies the requirements for this mechanism to operate efficiently and discusses 
the resultant limits on the achievable maximum energies when applied to AGN-type jets.

\section{Particle Spectra}
\begin{figure}[htbp]
\begin{center}
\includegraphics[width = 0.47 \textwidth]{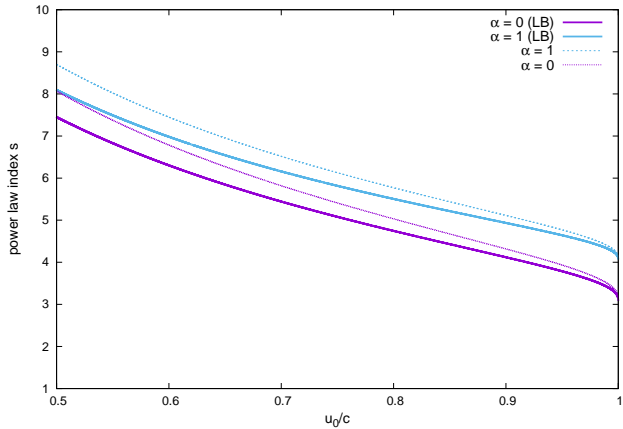}
\caption{Power-law index $s$ for the particle distribution $f(p)\propto p^{-s}$ as a function of the on-axis 
velocity $u_0$. A linearly decreasing velocity profile $u(r)$ with $u_0$ at $r_0$ and $u_2=0$ at $r_2$ has 
been assumed. The curves show the evolution in the assumed leaky box (LB) approach for a scattering 
time $\tau_s \propto p^{\alpha}$ with $\alpha=0$ and $1$, respectively. For comparison, results of analytical 
solutions of the full particle transport equation are shown as thin (dotted and dashed) lines.}
\label{fig1}
\end{center}
\end{figure}

As a stochastic particle acceleration process, the space-independent part of gradual shear acceleration obeys a 
diffusion equation in momentum space \citep[e.g.,][]{Earl1988,Rieger2006}. While moving across the velocity shear 
the particle momentum relative to the flow changes, so that in the local scattering frame a net increase in momentum 
can occur. Hence particle acceleration is closely tied to the diffusive transport across the flow. This, however, also 
implies that particles can diffusively escape from the system, impacting on the shapes of possible particle spectra. 
This becomes particularly relevant for non-relativistic flow speeds where cross-field escape counterbalances efficient 
particle acceleration. When diffusive escape and radiative losses are neglected, non-relativistic gradual shear 
acceleration is known to lead to power-law particle spectra $n(p) \propto p^2 f(p) \propto p^{-(1+\alpha)}$ for an 
energetic particle diffusion coefficient scaling as $\kappa \propto p^{\alpha}$ \citep{Berezhko1982,Rieger2006}. 
With reference to an analytical steady state model based on the full particle transport equation, \citet{Webb2018,
Webb2019} on the other hand recently showed that such hard power-law spectra are only achieved in relativistic 
shear flows, while the expected spectra become significantly softer for non-relativistic flow speeds. The present 
paper aims to explore and recapture this by means of a simple analysis.\\
Starting point is the standard momentum-space diffusion equation with spatial escape incorporated by means of 
a simple momentum-dependent particle escape term $f/\tau_{\rm esc}(p)$, i.e.,
\beq\label{diffusion}
\frac{\partial f}{\partial t} = \frac{1}{p^2} \frac{\partial}{\partial p }\left(p^2 D_p \frac{\partial f}{\partial p} \right) 
                                       - \frac{f}{\tau_{\rm esc}}\,. 
\eeq Here, $D_p$ denotes the momentum-space shear diffusion coefficient given by \citep{Rieger2006}
\beq
  D_p = \Gamma p^2 \tau_s \propto p^{2+\alpha}\,,
\eeq where $\tau_s(p)$ is the (momentum-dependent) mean scattering time assumed to follow a parameterization
$\tau_s(p) = \tau_0~(p/p_0)^{\alpha}$. $\Gamma$ denotes the shear coefficient. For a simple shear flow velocity 
profile $\vec{u} = u_z(r) \vec{e}_z$, appropriate for a cylindrical outflow, one finds \citep{Rieger2004,Webb2018}
\beq
 \Gamma = \frac{1}{15}\, \gamma_b(r)^4 \left(\frac{\partial u_z}{\partial r} \right)^2\,,
\eeq where $\gamma_b(r) =1/(1-u_z^2(r)/c^2)^{1/2}$.
Following eq.~(\ref{diffusion}) the characteristic (co-moving) particle acceleration time scale can be expressed as 
\citep[e.g.,][]{Rieger2019}
\beq\label{tacc}
 t_{\rm acc}(p) = \frac{c}{(4+\alpha)\, \Gamma \,\lambda} \propto p^{-\alpha}\,,
\eeq where $\lambda(p) = \tau_s(p) c$ is the particle mean free path. The typical escape time, on the other hand, 
is determined by cross-field transport, i.e. 
\beq 
 \tau_{\rm esc}(p) \simeq \frac{(\Delta r)^2}{2\,\kappa(p)}\propto p^{-\alpha} \,,
\eeq where $\kappa(p)=\lambda(p) c/3$ is the spatial diffusion coefficient and $\Delta r$ the width of the velocity shear 
region. Note that $t_{\rm acc}$ and $\tau_{\rm esc}$ have the same momentum-dependence.\\
The approach in eq.~(\ref{diffusion}) is somewhat analogous to the leaky-box model used to describe cosmic 
ray transport in the Galaxy, in which spatial diffusion and convection is replaced by an escape term. In such models 
particles are considered to propagate freely with a small probability ($1/\tau_{\rm esc}$) of escape each time they reach 
the boundaries. The probability of a particle remaining in the box then is $\exp(-t/\tau_{\rm esc})$, and the particle 
distribution inside the containment region is uniform.\\
Looking for steady-state solutions of eq.~(\ref{diffusion}) and employing a power-law Ansatz 
\beq
 f(p) = f_0\, p^{-s},
\eeq the power-law index above injection $p_0$ is given by
\beq
 s = \frac{(3+\alpha)}{2} + \sqrt{\frac{(3+\alpha)^2}{4} + (4+\alpha)\, \frac{t_{\rm acc}}{\tau_{\rm esc}}}\,.
 \eeq Consequently, only for $t_{\rm acc} \ll \tau_{\rm esc}$, i.e. only for fast shear flows with $(\partial u_z/\partial r)
 (\Delta r) \rightarrow c$, is the power law $f(p) \propto p^{-(3+\alpha)}$ recovered, in which case the exponent only 
 depends on the momentum dependence of the diffusion coefficient.\\ 
 For illustration, consider a linearly decreasing velocity profile $u_z(r) = u_0 - (\Delta u_z/\Delta r)\, (r-r_0)$ with 
 $\Delta u_z/\Delta r  = (u_0-u_2)/(r_2-r_0)$, where the subscript $2$ refers to quantities at the outer shear 
 boundary, and where for the following we assume $u_2=0$ and $r_0=0$. Then, $(\partial u_z/\partial r) = (\Delta u_z/
 \Delta r)$, and formally
 \beq
 \frac{t_{\rm acc}}{\tau_{\rm esc}} = \frac{10}{(4+\alpha) \gamma_b(r)^4 \left(\frac{\Delta u_z}{c}\right)^2}\,.
 \eeq To treat the $r$-dependence in this expression, noting the second-order dependence on the velocity gradient, 
 we replace $\gamma_b(r)^4$ by $\langle \gamma_b(r)^2\rangle^2$, where $\langle \rangle$ denotes averaging 
 over $r$. This yields
 \beq
 s  =  \frac{(3+\alpha)}{2} + \sqrt{\frac{(3+\alpha)^2}{4} + 40\,\left(\ln\frac{(1+u_0/c)}{(1-u_0/c)}\right)^{-2}}\,.
\eeq The evolution of the power-law index $s$ as a function of $u_0$ is shown in Fig.~\ref{fig1}. Obviously, for
non-relativistic flow speeds $u_0$ the spectra can be much steeper, approaching the limiting value $s=(3+\alpha)$ 
only at relativistic speeds $u_0 \rightarrow c$. 
In Fig.~\ref{fig1} we also show the evolution of the power-law index based on analytical solutions $f(r,p)$ of the 
full particle transport equation \citep{Webb2018} for comparison. No one-to-one correspondence is expected, 
though, as a specific $r$-dependence of the scattering time $\tau_s(r,p)$ has been assumed in the derivation of 
these solutions (typically resulting in $\tau_s \rightarrow \infty$ as $r \rightarrow 0$), and as the leaky-box 
approach implies a simplified treatment of spatial diffusion. Nevertheless, the qualitative behaviour is reasonably 
well reproduced, deviations being at the $\sim10\%$ level. For $\gamma_b(r_0)=4$ for example, one obtains 
$s=3.6$ ($\alpha=0$) and $s=4.5$ ($\alpha=1$), respectively. Note, however, that the expected power-law index 
$s$ is in general sensitive to the employed velocity profile, with steeper shapes towards lower speeds being 
possible \citep[e.g.,][]{Webb2019}. However, this effect becomes less important in the relativistic limit and 
a detailed analysis is left to a future paper.\\
The results shown here nicely illustrate that efficient shear acceleration requires relativistic velocity gradients. In 
principle such velocity gradients appear to be possible in AGN, not only on smaller (sub-parsec) but also on larger 
(kilo-parsec) jet scales, in particular in view of recent jet simulations showing that backflow speeds in AGN can 
be substantial \citep[e.g.,][]{Perucho2007,Rossi2008,Matthews2019,Perucho2019b}. We note that even for a less 
powerful FR I jet source such as M87, superluminal motion has been seen on kpc-scales \citep[e.g.,][]{Meyer2017,
Snios2019}. \\

\section{Maximum Energies}
While experiencing shear acceleration, particles can also lose energy via synchrotron radiation on a characteristic 
(comoving) timescale $t_{\rm syn}=\frac{9m^3c^5}{4e^4\gamma B^2}$. This becomes particularly relevant for 
electrons. One can estimate achievable maximum energies ($\gamma_{\rm max}$) by equating the acceleration 
timescale (cf. eq.~[\ref{tacc}]) with the loss timescale. For simplicity we consider a quasi-linear type parameterisation 
for the particle mean free path \citep[e.g.,][]{Liu2017} in the following, i.e. 
\beq
\lambda\simeq  \xi^{-1} r_g\left(\frac{r_g}{\Lambda_{\rm max}}\right)^{1-q} \propto \gamma^{2-q}\,, 
\eeq where $\xi \leq 1$ denotes the energy density ratio of turbulent versus regular magnetic field $B$, 
$\Lambda_{\rm max}$ is the longest interacting wavelength of the turbulence, $r_g$ is the particle Larmor 
radius, $\gamma$ the particle Lorentz factor, and $q$ is the power index of the turbulence spectrum (i.e.,
$q=1$  for Bohm-, $q=3/2$ for Kraichnan-, and $q=5/3$  for Kolmogorov-type turbulence). In our notation, 
$\alpha = 2-q$. Hence, for $0<\alpha<1$ we obtain 
\beq\label{gamma_max}
\gamma_{\rm max}  = \left[\frac{9\,(4+\alpha)(m c^2)^{3+\alpha} (\Gamma/c^2) \,
                                  \Lambda_{\rm max}^{1-\alpha} }{4\, \xi \,e^{4+\alpha} B^{2+\alpha}}\right]^{\frac{1}{1-\alpha}}\,,
\eeq with $\gamma_{\rm max}$, $\gamma$ and $B$ measured in the comoving frame. For $\alpha>1$, on
the other hand, acceleration, once operative, proceeds faster than synchrotron cooling. For a Kolomogorov-type 
turbulence ($\alpha =1/3$) and the linearly decreasing flow profile above with $\gamma_b(r_0)=3$, $\xi=0.2$ 
and $\Lambda_{\rm max} = \Delta r$, where $\Delta r$ is the lateral width of the shear layer, eq.~(\ref{gamma_max}) 
evaluates to
\beq\label{gamma_max_e}
 \gamma_{\rm e, max} \simeq 3.5 \times10^8\,\left( \frac{30~\mu \mathrm{G}}{B}\right)^{7/2} 
                                                    \left(\frac{0.1~\mathrm{kpc}}{\Delta r}\right)^2\,,
\eeq suggesting that electron Lorentz factors $\gamma_e \sim (10^8-10^9)$ are in principle achievable in the
large-scale jets of AGN. This would provide support to the electron synchrotron interpretation of extended X-ray 
emission in AGN jets that requires to sustain ultra-relativistic electrons along the jet \citep[e.g.,][]{Harris2006,
Georg2016}. 
Note that for a Kraichnan-type turbulence ($\alpha=1/2$), the numerical value, eq.~(\ref{gamma_max_e}), 
would be reduced by a factor of $\sim20$. In principle, due to the inverse dependence of $t_{\rm acc}$ on $\gamma$ 
(eq.~[\ref{tacc}]) efficient electron acceleration typically requires the injection of energetic seed particles. The 
latter could, however, most likely be provided by conventional Fermi-type acceleration processes 
\citep{Liu2017,Rieger2019}.

On the other hand, given their larger mean free path, shear acceleration of hadronic cosmic rays (CRs) is 
usually much easier to achieve. This could be of relevance for the origin of the highest energy CRs. Current 
evidence suggests that the CR composition around $10^{18}$ eV is dominated by light primaries. Given the 
observed level of isotropy in arrival directions, these CRs have to be of extragalactic (possibly AGN-type) 
origin so as to avoid a large anisotropy towards the Galactic Plane. With increasing energy the composition 
then seems to become more heavier ($\log(E_t$[eV])=18.3 transition), with a trend that protons are gradually 
replaced by helium, helium by nitrogen etc, an iron contribution possibly emerging above log(E[eV]) = 19.4 
\citep[e.g., see][for reviews]{Alves_Batista2019,Kachelriess2019}.

\begin{figure}[htbp]
\begin{center}
\includegraphics[width = 0.47 \textwidth]{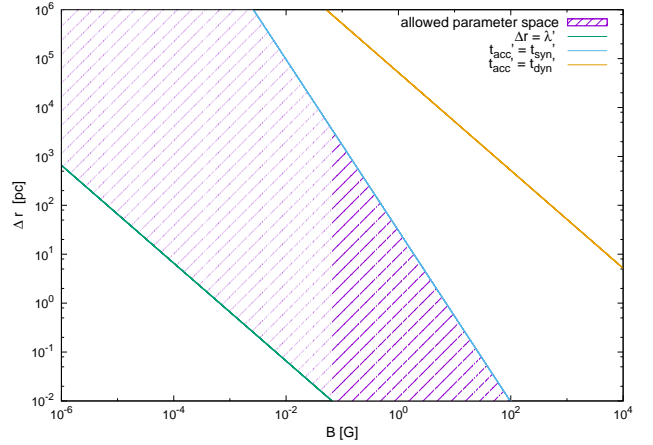}
\caption{Allowed parameter range (shaded) for shear acceleration of CR protons to energies $E_p'=10^{18}$ 
eV for a particle mean free path $\lambda'  \propto p'^{\alpha}$ with $\alpha=1/3$ (corresponding to Kolmogorov
type turbulence $q=5/3$). A flow Lorentz factor $\gamma_b(r_0)=3$ has been assumed.} 
\label{fig2}
\end{center}
\end{figure}

To enable shear acceleration of cosmic-ray protons to energies $E_t$ in the laboratory frame, corresponding to 
$E_t'=E_t/\gamma_b$ in the comoving frame, CR particles need to satisfy \citep[cf.,][]{Liu2017,Webb2019} (i) 
the (lateral) confinement condition, $\lambda(E_t') \leq \Delta r$,  (ii) the efficiency condition 
$t_{\rm acc} \leq t_{\rm syn}$ and (iii) the longitudinal confinement constraint $t_{\rm acc} \leq t_{\rm dyn} = d/(u_z
\gamma_b)$, where $d$ is the jet length, and $t_{\rm dyn}, t_{\rm acc}, t_{\rm syn}$ refer to the comoving frame. 
Figure~\ref{fig2} shows the parameter space (shear layer width $\Delta r$ versus comoving magnetic field strength 
$B$) permitted by these constraints in the case of $\alpha=1/3$ for the linearly decreasing shear flow profile above
with $\gamma_b(r_0)=3$. A jet shear width-to-length ratio $\rho_w = \Delta r/d=0.02$, and $\xi=1$ has been 
assumed in these calculations. 
For a magnetic field strength of $10^{-5}$ G for example, a width $\gppr 0.1$ kpc would be required. Such conditions 
are likely to be satisfied in the large-scale jets of AGN. Inspection of Fig.~\ref{fig2} indicates that for a given
jet width and magnetic field, the Hillas-type \citep{Hillas1984} confinement condition (i) usually imposes the tightest
constraint on the maximum CR energy. This suggests that CR particles are able to reach
\beq
 E_{\rm CR}' \simeq 3\times 10^{18} Z\, \xi^{\frac{1}{\alpha}} \left( \frac{B}{30~\mu \mathrm{G}} \right) 
                                                    \left( \frac{\Delta r}{0.1~\mathrm{kpc}} \right)\;\mathrm{eV}\,,
\eeq where $Z$ is the charge number. In the case of strong turbulence, $\xi \sim 1$, proton acceleration to
$E_t$ appears feasible, with the composition gradually becoming heavier. Note that due to the inverse scaling 
$t_{\rm acc} \propto 1/\lambda$ efficient injection may take place at different energy thresholds, and detailed 
modelling would be required to estimate the relative CR contribution at the highest energies. Pick-up shear 
acceleration of PeV CR protons (similar to our own Galaxy), however, is possible in the case of
$\alpha=1/3$ as $t_{\rm acc}/t_{\rm dyn} \simeq 2 \times 10^{-3} (100\,\mathrm{kpc}/d)$.  The particle 
spectrum of cosmic rays escaping the acceleration region approximately follows $\dot{n}_{\rm esc}(p) \propto 
p^2 f(p)/\tau_{\rm esc} \propto p^{2+\alpha-s}$, and can thus be quite hard.
We note that the present approach is complementary to the non-gradual ones discussed in \citet{Kimura2018} 
and \citet{Caprioli2015}, which are applicable for sufficiently narrow layers and ultra-fast ($\gamma_b\sim 30$)
flow speeds, respectively \citep[e.g., see][for discussion]{Rieger2019}.

\section{Conclusion}
As shown here, fast shear flows can facilitate a continued Fermi-type acceleration of charged particles, 
capable of producing power law particle momentum distributions as long as the velocity shear persists. 
In general, however, relativistic flow velocities are required for this mechanism to operate efficiently. As 
discussed here, such velocities may be encountered in the jets of AGN. Evaluating achievable electron 
energies (synchrotron-limited to PeV [$10^{15}$ eV] energies) suggest that gradual shear acceleration 
could offer an interesting explanation for the extended high-energy emission observed in large-scale 
AGN jets. Similarly, EeV [$10^{18}$ eV] energies (confinement-limited) may be achieved for cosmic-ray 
protons, indicating that shear acceleration in AGN jets could play a relevant role in the energisation of 
the observed ultra-high energy cosmic rays. While these estimates are based on a simplified treatment and 
more extended studies are required, it seems hard to see, how velocity shear could not play a role in the 
energization of charged particles. 

\acknowledgments
FMR kindly acknowledges funding by a DFG Heisenberg Fellowship RI 1187/6-1. Discussions 
with Gary Webb and Martin Lemoine are gratefully acknowledged. We thank the anonymous 
referee for insightful comments.


\bibliography{references}{}
\bibliographystyle{aasjournal}



\end{document}